# Enhanced backscatter of optical beams reflected in turbulent air

W. Nelson[1,*], J.P. Palastro[2], C. Wu[1], and C.C. Davis[1]

[1]Department of Electrical and Computer Engineering, University of Maryland, College Park, Maryland 20740
[2]Naval Research Laboratory, Washington, DC 20375-5346, USA
*Corresponding author: wnelson2@umd.edu

Optical beams propagating through air acquire phase distortions from turbulent fluctuations in the refractive index. While these distortions are usually deleterious to propagation, beams reflected in a turbulent medium can undergo a local recovery of spatial coherence and intensity enhancement referred to as enhanced backscatter (EBS). Here we validate the commonly used phasescreen simulation with experimental results obtained from lab-scale experiments. We also verify theoretical predictions of the dependence of the turbulence strength on EBS. Finally, we present a novel algorithm called the "tilt-shift method" which allows detection of EBS in frozen turbulence, reducing the time required to detect the EBS signal.

## 1. Introduction

Optical beams propagating through the atmosphere acquire phase distortions from turbulent fluctuations in the refractive index. These distortions result in spreading, wander, and scintillation of the beam, reducing the accuracy and efficiency of applications requiring long range propagation, such as free space optical (FSO) communications and directed energy (DE) laser applications [1-5]. Fortunately, not all phenomena associated with propagation through atmospheric turbulence are detrimental. When an optical beam propagates through turbulence to a target, reflects off the target, and propagates back through the same turbulence, it undergoes a local recovery of spatial coherence and intensity enhancement referred to as enhanced backscatter (EBS) [4-16]. EBS is observed as a retroreflected, localized intensity spot in the detector or receiver plane when many independent instances of turbulence are averaged over.

EBS was extensively studied after the first theoretical predictions of the phenomenon in 1972 by Belenkii and Mironov [17]. Numerical investigation followed shortly after using the phase screen model to represent turbulence induced phase distortions [11]. EBS was later observed in lab experiments [12] and field experiments [13] in 1977 and 1983 respectively. Further studies have demonstrated the potential of EBS for precision pointing and tracking [15]. However, there have been fewer recent investigations, possibly due to minimal applications and detection difficulty, but major advances in both computing and imaging technologies warrant further studies of EBS.

Our objective here is fivefold: First to experimentally measure EBS from retroreflectors and diffuse reflectors using modern imaging systems. Second, to validate propagation simulations of EBS with experimental results. Third, to demonstrate a scaling between the experiments and DE-scales, justifying use of the simulations for examining EBS in this regime. Fourth, to verify theoretical predictions on the influence of turbulence strength on EBS. And finally, to present a novel method for detecting EBS, which greatly reduces the time required for observation.

We begin by defining terms used throughout the paper. We define the transmitter, target, and receiver planes as the planes where the beam starts, reflects, and ends respectively. For our purposes the transmitter and receiver planes are the same. A double pass refers to the beam propagating from the transmitter plane to target plane and back to the receiver plane. A bistatic channel refers to a channel where the turbulence on the path to the target is uncorrelated with the turbulence of the return path. Bistatic channels arise when the reflected beam path does not overlap with the initial beam path or when the turbulence correlation time is shorter than the double pass transit time. A monostatic channel refers to a channel where the turbulence is identical on the path to the target and the return path. Monostatic channels arise when a beam is reflected at a small angle back toward the transmitter and the turbulence correlation time is longer than the double pass transit time.

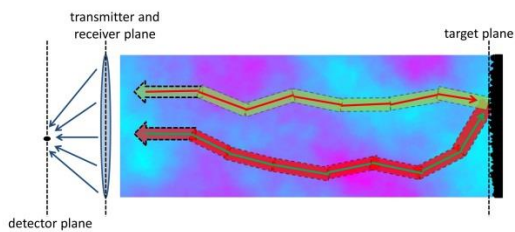

Figure 1. Diagram of a pair of reciprocal rays taking the same path in opposite directions.

EBS can be understood through a ray model of a collimated beam propagating through a turbulent channel and reflecting from a diffuse reflector. In the ray model, a collimated laser beam consists of many individual rays starting with zero transverse wavenumber in the transmitter plane. As the rays propagate through the atmosphere, the turbulent fluctuations in the refractive index change each ray's transverse wavenumber. When the rays reach the target plane they are scattered back toward the transmitter plane along different paths than those taken to the target. A subset of the rays will end with zero transverse wavenumber in the transmitter plane and within the aperture of the outgoing beam. If the channel is monostatic, because these rays end up within the aperture of the outgoing beam, there must exist a ray that started at the same location and took the same path through the turbulence but in the opposite direction. These rays are referred to as reciprocal rays and always come in pairs. Reciprocal rays pass through the same turbulence cells but in reverse order. Since their path length is the same, the two rays forming a reciprocal ray pair are always in phase at the receiver plane [6]. The reciprocal rays will be normally incident on a lens placed in the receiver plane, and refracted to the focal point. While a reciprocal ray pair is always in phase, each pair is generally out of phase with all other reciprocal pairs: each pair has a different path length through the turbulence. This leads to high average intensity and a high variance in intensity at the focal point of the lens, known as enhanced backscatter. The intensity at the focal point is in general predicted to be increased by a factor of two [6,14]. The ray model gives a concise, visual overview of EBS, yet provides a remarkable level of insight into the phenomenon [6,14,15]. For a comprehensive wave theory derivation of EBS see [7] and [9].

EBS is a consequence of the reciprocity of atmospheric turbulence [18]. Reciprocity, or the symmetry with respect to interchange of the source and receiver locations [19], is an important property with powerful implications. Using reciprocity, Lukin and Charnotskii showed that the field from a point source on a target in the detector plane is proportional to the field of a beam on a target [20,21]. In a monostatic turbulent channel, the point in the detector plane that possesses this property coincides with the location of the EBS peak. Thus observation of the EBS peak reveals a point at which, in theory, instantaneous measurements of the beam intensity on target can be made [20]. Furthermore, observation of EBS ensures that the atmospheric channel is reciprocal, which is not true when long distances or abnormally high wind velocities are considered. This is our motivation to investigate EBS.

Here we examine EBS on two length scales. The smaller scale corresponds to lab experiments using retroreflector and diffuse reflector targets in which fan assisted hot plates generate strong turbulence over a short distance. We show that the standard simulation model for propagation through turbulence, a steady-state paraxial wave equation with optical turbulence modeled with phase screens [2,4,22], largely agrees with experimental observations of the EBS phenomenon. We then use the simulation to consider a second, larger length scale relevant to DE applications. At these length scales, we find that EBS detection requires weak and strong levels of turbulence for retroreflectors and diffuse reflectors, respectively. We also find that although the EBS phenomenon is robust at this larger scale, the standard method of averaging over independent instances of turbulence is not practical. Performing a temporal average over turbulence causes two problems: a delay in detecting the EBS signal, and when the EBS signal is detected, it does not represent the current state of the channel. We instead introduce a novel method of EBS detection called the tilt-shift method (TSM), in which a spatial average replaces the standard temporal average.

We begin in section 2 by describing the experimental arrangement for measuring the intensity profile of a laser beam propagated through lab-scale, hotplate generated turbulence. We then verify the propagation simulation reproduces the experimental results. In section 3, we scale the simulation to parameters relevant to DE applications and present results comparing the EBS signal from retroreflectors and diffuse surfaces. Section 4 describes the TSM which enables the detection of EBS in frozen turbulence. Section 5 ends the manuscript with a summary of our results and outlook for the future.

## 2. Lab-scale turbulence

### A. Experimental arrangement

A schematic of the experimental arrangement is shown in figure 2 for both the monostatic and bistatic cases. An initially collimated beam with wavelength $\lambda = 635$ nm and spot size $w_0 = 1$ cm originates from a Thor Labs fiber laser. The beam propagates through hotplate generated turbulence, reflects off the target then propagates back through the turbulence once more before being focused onto the CCD camera. The same optical components are present in the bistatic and monostatic channels so that any loss or deformations to the beam from the components is the same for each channel.

The turbulent channel is 3 m long and fueled by two $60 \text{ cm} \times 7.5 \text{ cm}$ hotplates that can reach up to 200°C. Through thermal conduction the hotplates generate a turbulent boundary layer. The addition of a duct fan provides a jet stream down the axis of the channel that facilitates energy cascade and sustains the turbulent flow. The hotplates are wide enough to allow a slant pass across the channel for the transmitted beam in the bistatic arrangement.

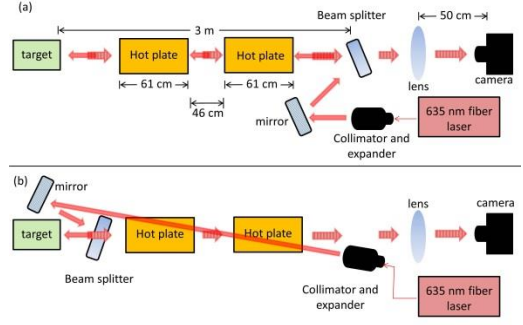

Figure 2. Schematic of the experimental arrangement for the (a) monostatic and (b) bistatic channels

The experiment was conducted with two types of target: a retro and a diffuse reflector. A 2 inch diameter dielectric cornercube was used as the retroreflector, and a simple white poster board for the diffuse reflector. The reflected beam passed through a 50/50 beam splitter and was focused with a 50 cm focal length, 2 inch lens (5.08 cm) onto a CCD camera with $1024 \times 1024$ square pixels 5.5 μm on a side placed in the focal plane. The exposure time of the camera was set to $30\,\mu s$ and 15 ms for the retroreflector and diffuse reflector respectively. The intensity was averaged over 100 frames collected at 8 frames per second.

### B. Simulation description

Following our previous work [22], we use the steady-state paraxial wave equation with optical turbulence modeled with phasescreens to solve for the transverse electric field of a laser beam propagating through the turbulent channel. For a detailed description of the algorithm see reference 22. The transverse simulation domain is 204.8 mm × 204.8 mm with a resolution of .1 mm for the retroreflector, and 51.2 mm × 51.2 mm with a resolution of .025 mm for the diffuse reflector. The turbulence is characterized by an analytic approximation to the Hill spectrum [4]

$$\Phi_n(\kappa) = 0.033 C_n^2 \left[1 + 1.802\left(\frac{\kappa}{\kappa_l}\right) - 0.254\left(\frac{\kappa}{\kappa_l}\right)^{7/6}\right] \frac{\exp\left(-\kappa^2/\kappa_l^2\right)}{\left(\kappa^2 + \kappa_0^2\right)^{11/6}} \quad (1)$$

where $\kappa_l = 3.3/l_0$ and $\kappa_0 = 2\pi/L_0$. Following the method described in [23], we experimentally measured the scintillation index and mean squared angle of arrival to be $\sigma_I^2 = (6.3 \pm .2) \times 10^{-4}$ and $\langle \alpha^2 \rangle = (1.4 \pm .1) \times 10^{-8}$ rad$^2$ respectively. Using these values along with the equations [24]:

$$\langle \alpha^2 \rangle = 3.28 C_n^2 L l_0^{-1/3} \quad (2a)$$
$$\sigma_I^2 = 12.8 C_n^2 L^3 l_0^{-7/3} \quad (2b)$$

we calculate the inner scale and effective $C_n^2$ associated with our hotplate generated turbulence to be $l_0 = 8.7 \pm .5$ cm, and $C_{n,eff}^2 = (6.2 \pm .7) \times 10^{-10}$ m$^{-2/3}$ respectively. The outer scale is chosen to be $L_0 = 60$ cm corresponding to the length of one hotplate. Fluctuations larger than the simulation domain are not accurately sampled. However, we are primarily concerned with small scale fluctuations that lead to scintillation so this does not present an issue. We note that this could be resolved through the addition of sub-harmonics in the process of generating phasescreens [25].

A double pass in a bistatic turbulent channel uses eight phasescreens: four for propagation from the transmitter plane to the target plane, and four for propagation from the target plane to the receiver plane. A double pass in a monostatic turbulent channel uses four phasescreens: the same phasescreens are used for propagation in both directions. The m$^{th}$ phasescreen is placed at a distance $(m-1/2)L/n$ where n is the number of phasescreens and L is the distance from the transmitter to the target. We model the propagation from the receiver plane to the detector plane, the focal plane of the lens, by taking the Fourier transform of the electric field in the receiver plane and scaling the domain lengths to correspond to focusing by a 50 cm focal length lens. All plotted quantities are considered in the detector plane. Intensity averages are performed over 1000 runs through statistically independent turbulent channels.

### C. Retroreflector

First we examine the experimental and simulation results for the cornercube target. The cornercube inverts and reverts the wavefront corresponding to a 180 degree rotation and a reversal in the direction of propagation. This results in many reciprocal ray pairs leading to a distinctive EBS intensity peak in the detector plane. Figures 3a and 3b display the average intensity profiles for the monostatic and bistatic channels respectively obtained using the experimental arrangements described in section 2A. Figure 3c compares the y = 0, x-slice for both channels. Intensity is normalized according to the equation $I_{norm}(x,y) = \langle I(x,y) \rangle / \langle I_{bi}(0,0) \rangle$ where $I_{bi}(0,0)$ is the intensity at the center of the bistatic beam. The average intensity in the monostatic channel exhibits the sharp, central peak reaching a value of ~2 indicative of an EBS enhancement . Other than the EBS enhancement, the monostatic and bistatic channels produce nearly identical Gaussian intensity profiles.

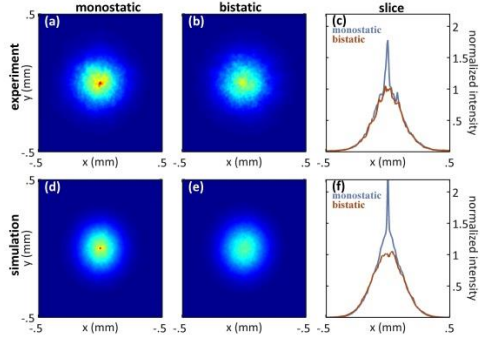

Figure 3. Average monostatic (first column), bistatic (second column) and x-slice (third column) intensity profiles from a retroreflector target obtained from the experiment (first row) and simulation (second row).

In the simulation, the retroreflection at the target plane is applied by rotating the propagation domain 180 degrees and propagating the beam in the opposite direction. Figures 3d, 3e, and 3f display the average intensity profiles obtained from the simulation for the monostatic channel, the bistatic channel, and the y = 0, x-slice for both channels respectively. The turbulent beam spreading in the simulation matches the experiment, indicating that the strengths of turbulence are approximately equal. The primary difference in the intensity profiles is the sharper contrast of the EBS peak to the background predicted by the simulation. This could be due to small errors in alignment or aberrations from the optical components in the experimental arrangement, as the simulation assumes perfect alignment and flawless optical components. Regardless of this small difference, the 2-dimensional phasescreen simulation accurately models the 3-dimensional turbulence of the experiment and reproduces the EBS observed from a retroreflector.

### D. Diffuse reflector

We now examine the experimental and simulation results for a diffuse reflector. In the experiment a large white poster board provides the diffuse reflection. Figures 4a, 4b, and 4c display the observed average intensity profiles for the monostatic channel, the bistatic channel, and the y = 0, x-slice for both channels respectively. As in section 2C, the intensity is normalized to the value at the center of the bistatic beam. The reflection from the rough surface creates a broad speckle pattern in the intensity profile for both the monostatic and bistatic channel. Again the monostatic channel produces an EBS peak with twice the bistatic intensity, but the peak is much wider than that resulting from the cornercube target.

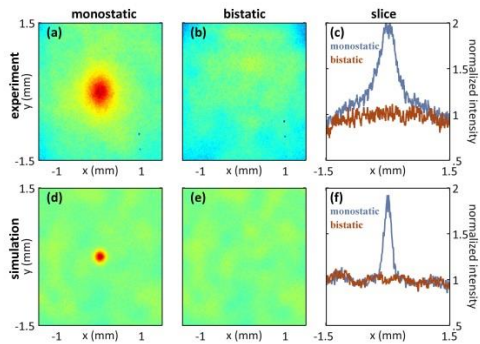

Figure 4. Average monostatic (first column), bistatic (second column) and x-slice (third column) intensity profiles from a diffuse reflector target obtained from the experiment (first row) and simulation (second row).

A diffuse reflection is difficult to model precisely due to both the random and fractal nature of the surface when viewed on the scale of optical wavelengths. The surface scatters the beam to large angles contributing large transverse wavenumbers, $|\mathbf{k}_\perp|$, to the electric field, which, in turn, cause significant beam spreading. Thus it becomes numerically intractable to make the Fourier and real space domains large enough to both resolve the large $|\mathbf{k}_\perp|$ values and follow the beam spreading. However, we are able to ignore $|\mathbf{k}_\perp|$ values larger than our Fourier domain for two reasons. First, the resulting reduction in beam power does not significantly affect the average intensity ratio of beams that propagate through different channels but are reflected from the same surface. Second, the small scale phase distortions occur at random uncorrelated points and their associated $|\mathbf{k}_\perp|$'s leave the propagation domain of interest upon diffraction.

To simulate a reflection from a diffuse surface, we first apply a random phase uniformly distributed from zero to $2\pi$ to the electric field envelope at each grid point [26]. The random phases are pairwise uncorrelated. We then apply a filter in the Fourier domain that truncates $|\mathbf{k}_\perp| > 1.1 \times 10^5$ rad/m to ensure that all directions are limited by the same $|\mathbf{k}_\perp|$. This step is required because the simulation domain is square. We note that if the reflection model creates a $|\mathbf{k}_\perp|$ that is larger than our Fourier domain, it will be aliased to another location within our Fourier domain—a property of discrete Fourier transforms. Normally this would corrupt a Fourier transform, but since the phase distortions from the diffuse reflection are random and uncorrelated, the aliasing results in a slightly different but equivalent random phase. Comparison of low-pass filters with different cutoff frequencies reveals that the cutoff frequency has no discernable impact on the EBS peak.

Figures 4d, 4e, and 4f display the average intensity profiles obtained from the simulation for the monostatic channel, the bistatic channel, and the y = 0, x-slice for both channels respectively. The monostatic and bistatic intensity profiles both display a speckle pattern characteristic of a diffuse reflection as displayed in figures 4a and 4b. The EBS intensity enhancement factor of 2 is observed in the monostatic channel simulation. There is, however, a disagreement in the width of the simulated and measured EBS peaks. In particular, the measured EBS peak is approximately twice as wide as the peak predicted by the simulations. This is most likely due to differences in the actual reflections from the white poster board target and those resulting from our diffuse reflector model. It has been shown that the roughness and material properties of the scattering surface impact the width of the EBS peak [27]. Since we are investigating EBS in the context of DE where we are generally unable to know the target's material properties, we are not concerned with the difference in EBS peak width.

## 3. Directed energy scale

### A. Transition to DE length scales

Using the Rytov variance and the Fried parameter [4]:

$$\sigma_R^2 = 1.23 C_n^2 k_0^{7/6} L^{11/6} \quad (3a)$$
$$r_0 = 1.67 (C_n^2 k_0^2 L)^{-3/5} \quad (3b)$$

as measures of the turbulence strength and transverse coherence length at DE scales, respectively, we can scale the simulation parameters for larger beams propagating over longer turbulent channels. This allows us to investigate EBS on length scales relevant to common DE parameters while avoiding the procedural difficulties of conducting experiments at this scale.

First we calculate the Rytov variance and transverse coherence length for the lab-scale turbulence. Although the lab-scale propagation distance is small and the inner scale is large, we are justified in using the Rytov variance equation because we measure the intensity profile in the focal plane where the scintillation is fully developed (ie. The focal plane represents the far field). Our parameters give the value $\sigma_R^2 = .83$. Due to the large inner scale and short propagation distance of the lab generated turbulence, we use a different form for the transverse coherence length which depends on the size of the inner scale [4]:

$$\rho_0 = \left(1.64 C_n^2 k_0^2 L l_0^{-1/3}\right)^{-1/2} . \quad (4)$$

We now demonstrate that a lab-scale and DE-scale simulation are nearly identical up to a scale length when the Rytov variance and the ratio of the transverse coherence length to initial beam radius are fixed. We define the beam size ratio as $\alpha \equiv w_{DE}/w_{lab}$ where $w_{DE}$ and $w_{lab}$ are the $1/e$ radii of the electric field for the DE scale and lab-scale respectively. Noting the relation between the Rytov variance and Fried parameter in equation 3:

$$\sigma_R^2 \sim \left(\frac{L}{r_0^2 k_0}\right)^{5/6}, \quad (5)$$

we obtain the following expressions for the equivalent length and $C_n^2$ of the DE-scale channel:

$$L = .3(\sigma_R^2)^{6/5}(\alpha \rho_0)^2 k_0 \quad (6a)$$

$$C_n^2 = 7.4(\sigma_R^2)^{-6/5}(\alpha \rho_0)^{-11/3} k_0^{-3} \quad (6b)$$

Using the values of the experiment $\sigma_R^2 = .83$, $\rho_0 = 1.2$ mm, $w_{lab} = 1$ cm, and $w_{DE} = 20$ cm, we find $L \simeq 1.4$ km and $C_n^2 \simeq 8 \times 10^{-15}$ m$^{-2/3}$ for the equivalent DE-scale channel. The inner scale for the DE channel is chosen to be 1 mm, a common value for atmospheric turbulence.

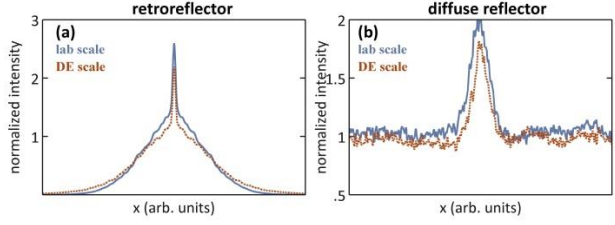

Figure 5. Comparison of lab and DE simulations for a retroreflector (a) and diffuse reflector (b)

Figures 5a and 5b compare the average monostatic intensity profiles of the lab-scale and equivalent DE-scale simulations for a retroreflector and diffuse reflector respectively. The x-axis of the large-scale intensity profile has been dilated by the factor $\alpha=20$. Although the inner scales differ by approximately two orders of magnitude, the average monostatic intensity profiles nearly match.

It is important to note that at the lab scale, both the beam width and final transverse coherence length are smaller than the inner scale. Scintillation develops when small and large scale phase fluctuations develop into amplitude fluctuations [28]. The small scale phase fluctuations arise from turbulent cells smaller than the transverse coherence length or beam width, whichever is smaller. However, the large inner scale of the laboratory generated turbulence greatly diminishes the amplitude of the small scale fluctuations. As a result, the scintillation observed in the focal plane results primarily from the lens-like, refractive phase perturbations imparted by the large scale turbulent fluctuations [28]. This is in contrast to the DE scale where both the large and small scale phase fluctuations contribute to scintillation. In spite of this difference between the lab and DE scale, Fig. 5 suggests that the EBS signal, in the far field, is relatively insensitive to the exact scale sizes of the initial phase perturbations.

## B. Directed energy parameters

Having compared the simulation data with experimental data at the lab scale, we now turn to parameters used in DE applications. We simulate the propagation of an initially collimated Gaussian beam with wavelength $\lambda=1\ \mu m$ and spot size $w_0=10\ cm$. The transverse simulation domain is 2.048 m by 2.048 m with a resolution of 1 mm. The distance from the transmitter plane to the target plane is 1 km. The turbulence is characterized by the Hill spectrum with $l_0=1\ mm$ and $L_0=1\ m$, which are common parameters for the inner and outer scale of atmospheric turbulence [4]. We consider $C_n^2$ values varying from $5\times10^{-16}\ m^{-2/3}$ to $5\times10^{-13}\ m^{-2/3}$ corresponding to weak and strong atmospheric turbulence respectively. All other parameters remain unchanged.

To quantify the intensity enhancement, it is convenient to define the enhancement factor, $F$, as

$$F = \frac{\overline{I_m}}{\overline{I_b}} \qquad (7)$$

where $\overline{I_m}$ is the average intensity at the focal point in the detector plane of a monostatic channel and $\overline{I_b}$ is the average intensity at the focal point in the detector plane of a bistatic channel. As previously explained, the expected value of the enhancement factor is two under conditions of sufficiently strong turbulence.

## C. Retroreflector

Figure 6a displays the value of the enhancement factor averaged over 400 independent turbulent channels as a function of $C_n^2$ and the corresponding Rytov variance for a corner cube target. Figure 6b displays the enhancement factor values over the first 200 runs at 6 different values of $C_n^2$. Figure 6 illustrates two major points: EBS requires averaging over many double passes, and strong turbulence is beneficial for EBS detection whereas it is usually a hindrance to long distance propagation.

At $C_n^2=1\times10^{-16}\ m^{-2/3}$ the turbulence is too weak for the EBS phenomenon to be easily observed and the enhancement factor is approximately 1. As $C_n^2$ increases to $5\times10^{-15}\ m^{-2/3}$ ($\sigma_R^2\approx.17$) the enhancement factor increases to 2. With a cornercube target, reciprocal paths exist even in the absence of turbulence due to the wavefront inversion in the target plane. Therefore the turbulence is not required to laterally shift rays to form reciprocal pairs and the turbulence only needs to be so strong as to produce minimal scattering effects. The relatively low levels of turbulence required for backscattered enhancement from a cornercube makes these targets suitable for experimental investigations of EBS as found in section 2.

Theoretical predictions of the enhancement factor dependence on turbulence strength divide the turbulence strength into three regimes: weak, strong, and saturated turbulence. EBS is first observed in the weak fluctuation regime, where the enhancement factor transitions from 1 to 2. In the strong fluctuation regime, the enhancement factor temporarily reaches a maximum of approximately 3. As the strength of turbulence increases into the saturated regime, the enhancement factor converges to 2 [9,10].

Figure 6a exhibits the same general behavior. The fluctuations in the enhancement factor are due to averages at each $C_n^2$ value being taken over 400 statistically independent simulations. The error in the enhancement factor, $\Delta F$, is estimated using propagation of errors where the error for the average monostatic and bistatic intensities are determined by their standard errors respectively. Specifically, we use $\Delta F = \pm(\bar{I}_b N^{1/2})^{-1}[\sigma_m^2 + \sigma_b^2(\bar{I}_m/\bar{I}_b)^2]^{1/2}$, where $\sigma_i^2$ is the variance for each channel. Strictly speaking, the result has not fully converged. However it is not difficult to identify the weak, strong, and saturated regimes as $\sigma_R^2 < .07$, $.07 < \sigma_R^2 < 1$, and $\sigma_R^2 > 1$ respectively.

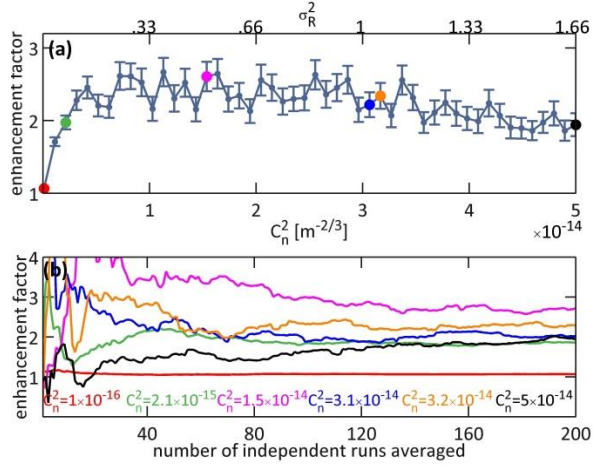

Figure 6. Enhancement factor as a function of $C_n^2$ (a) and number of independent runs averaged (b) for a retroreflector target.

Figure 6b confirms that the data in figure 6a was sufficiently averaged. It also illustrates the necessity to average over many instances of turbulence to obtain an accurate enhancement factor. In particular, the enhancement factor for $C_n^2 = 5 \times 10^{-14}$ temporarily drops below 1 even though the long term average is approximately 2. We note that these curves correspond to one specific set of statistically independent simulations, and a different set of runs with the same $C_n^2$ will average in a different manner but produce a similar result.

### D. Diffuse reflector

We repeat this investigation for a diffuse reflector with the same parameters. Unlike a retroreflector, a diffuse reflector in the absence of turbulence does not produce reciprocal ray pairs. Although some rays get reflected directly backward, these are not considered reciprocal ray pairs because each ray essentially pairs with itself. Therefore, any reciprocal ray path must be the result of turbulence induced scattering, and, as a result, a diffuse reflector requires stronger turbulence than a cornercube for EBS observation.

This is reflected in figure 7a which displays the value of the enhancement factor averaged over 400 independent turbulent channels as a function of $C_n^2$ and the corresponding Rytov variance. The enhancement factor first reaches a value of 2 at $C_n^2 = 4.2 \times 10^{-14}$ m$^{-2/3}$, $\sigma_R^2 = 1.4$ which is 20 times greater than the minimum value of $C_n^2$ required for the same enhancement from a retroreflector target. Correspondingly, the weak, strong, and saturated regimes are shifted to $\sigma_R^2 < 1.4$, $1.4 < \sigma_R^2 < 10$, and $\sigma_R^2 > 10$ respectively.

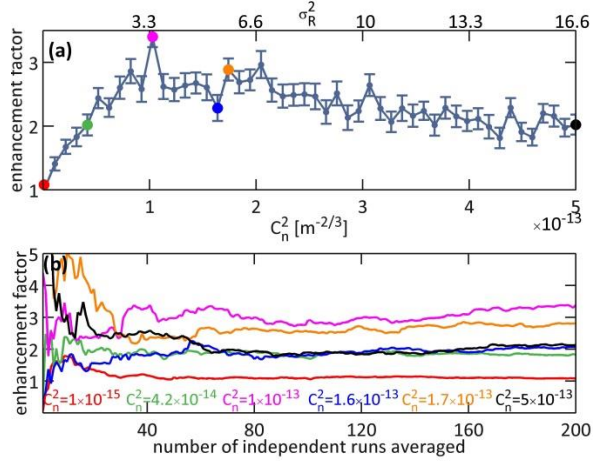

Figure 7. Enhancement factor as a function of $C_n^2$ (a) and number of independent runs averaged (b) for a diffuse reflector target.

Figure 7b displays the approach to convergence for the enhancement factor as a function of the number of independent runs averaged for 6 values of $C_n^2$. This figure illustrates an issue not encountered with a retroreflector. As indicated by figure 7a, turbulence characterized by $C_n^2 = 1\times 10^{-15}$ m$^{-2/3}$ is not within the EBS regime for a diffuse reflector. However figure 7b shows that the turbulence is strong enough to produce a transient enhancement factor, approximately 2, when too few runs are averaged over. This transient enhancement factor is the result of scintillation and not EBS. .

Atmospheric turbulence typically becomes uncorrelated over a time scale on the order of milliseconds [29] due to atmospheric conditions such as wind, temperature, and humidity. This means that intensity profiles should be sampled at a rate in the hundreds of Hz and it would take too long in the context of DE applications to distinguish an intensity enhancement due to EBS and an intensity enhancement due to scintillation.

## 4. Tilt-shift method

The standard method of detecting EBS requires averaging the intensity in the detector or receiver plane over many double passes through statistically independent turbulence. This puts a lower limit on the time delay between successive passes to ensure the turbulence has become uncorrelated. Furthermore, once the EBS signal has been obtained, it is only representative of the turbulence's average state not the current state. Here we introduce a novel method, which we refer to as the tilt shift method (TSM), that can detect EBS from a single instance of turbulence. As a result the method eliminates the time delay, and provides an EBS signal representative of the current state of the turbulence.

The method expands on the principle of reciprocity in the EBS phenomenon: the same distortions are applied to the beam on return path as the outgoing path. This implies we can modify the transmitted beam, as long as we make the same modification to the reflected beam. In the TSM, we apply a small angle tilt to the transmitted beam before it enters the turbulent channel. Instead of re-tilting the reflected beam after it exits the channel, we shift the coordinates in the detector plane according to the initial tilt. This is equivalent to re-tilting the beam because the detector plane is the focal plane of the lens. By shifting the coordinates, the reciprocal rays refract to the focal point while the majority of the reflected beam is diverted away from the focal point. We note that the coordinate shift can be accomplished by a processor and no extra hardware is required. Additional beams, each with a unique tilt, are propagated through the same channel, and given the appropriate shift in the detector plane. Each beam propagates along a different path and contributes a different set of reciprocal rays at the focal point. The EBS can then be observed in the intensity profile averaged over small angle tilts through the same turbulent channel. Said differently, the TSM replaces an average over paths through statistically independent channels for an average over different paths through the same channel. We note that the reflected beam path must overlap with the transmitted beam path, placing an upper limit on the tilt angle.

While the tilt shift method works for retroreflector and mirror targets, we choose to focus on a diffuse reflector because the wide angle of reflection increases the upper limit on the tilt angle. To simulate a tilt, we apply a complex phase to the initial field envelope of the i$^{th}$ beam, $E_{\perp,i}(\mathbf{r},0) \rightarrow E_{\perp,i}(\mathbf{r},0)e^{i\mathbf{k}_{tilt,i}\cdot\mathbf{r}}$ where the magnitude of the tilt wavenumber, $|\mathbf{k}_{tilt,i}| = k_{tilt} \cong k_0 \theta_{tilt}$, and absolute tilt angle, $\theta_{tilt}$, are equal for each beam. A unique tilt is defined through the angle $\phi_i$ where $k_{x,tilt,i} = k_{tilt}\cos\phi_i$ and $k_{y,tilt,i} = k_{tilt}\sin\phi$. The coordinates in the detector plane are shifted by $\mathbf{k}_{tilt,i}f/k_0$ where $f$ is the focal length of the lens.

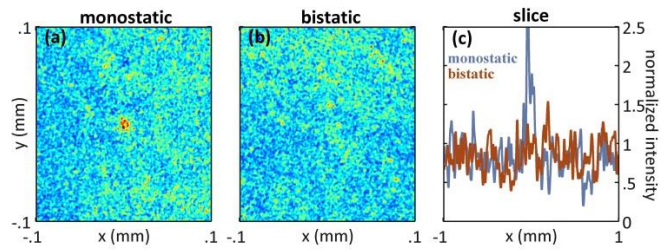

Figure 8. Average monostatic (a) bistatic (b) and x-slice (c) intensity profiles using the tilt shift method in static turbulence.

Figure 8 shows the monostatic and bistatic intensity profiles from the TSM using 16 equally spaced tilts ($\phi_{i+1} - \phi_i = \pi/8$) of magnitude $\theta_{tilt} = 52\ \mu rad$. The DE parameters described in section 3b are used along with $C_n^2 = 5 \times 10^{-14}\ \text{m}^{-2/3}$. Each beam propagates through the same frozen turbulent channel. We observe EBS in the monostatic channel from a single instance of turbulence.

## 5. Conclusion

We have observed EBS from both retroreflector and diffuse reflector targets. An effective value of $C_n^2$ and inner scale were determined for the lab turbulent channel.. These values were then used to verify that the phasescreen simulation agrees with the experimental measurements of EBS. We showed that the simulation can be scaled to length scales associated with DE applications while preserving the characteristics of EBS. Using the simulation, we investigated the dependence of turbulence strength on the enhancement factor. These results agreed with theoretical predictions. This investigation revealed that a diffuse surface requires much stronger turbulence than a retroreflector for observation of EBS. However, EBS is still a robust phenomenon for DE applications due to the long distances and high levels of turbulence commonly encountered. It also revealed that a rapid method of detecting EBS is required if EBS is to be used in DE systems. The TSM fulfills this requirement because it allows detection of EBS in frozen turbulence, thus eliminating the need to wait for the turbulence to change.

### Acknowledgments
The authors would like to thank L. Andrews, R. Phillips, P. Sprangle, and T. M. Antonsen Jr. for fruitful discussions. This work was supported by JTO through ONR under contract number N000141211029.